\documentclass[aip,reprint]{revtex4-1}

\usepackage[colorlinks=true,
            linkcolor=blue,
            citecolor=blue,
            urlcolor=blue]{hyperref}
\usepackage{graphicx}
\usepackage[utf8]{inputenc}
\usepackage[T1]{fontenc}
\usepackage{etoolbox}
\usepackage{lineno}
\usepackage[percent]{overpic}
\usepackage{float}
\usepackage{xcolor}
\usepackage{svg}
\usepackage{tikz}
\usepackage{amsmath}
\usepackage{amssymb}
\usepackage{orcidlink}
\usepackage{lineno}
\usetikzlibrary{decorations.pathmorphing}
\usepackage{xcolor}
\usepackage{soul} 
\usepackage{makecell}

\draft 

\makeatletter
\let\@fnsymbol\@fnsymbol@latex
\@booleanfalse\altaffilletter@sw
\def\frontmatter@makefnmark{%
  \@textsuperscript{\normalfont\@thefnmark}%
}
\makeatother

\begin{document}
\title{Exploring the proximity effect of thin tungsten-aluminum films for phonon-mediated kinetic inductance detectors}


\newcommand{\INFNRoma}{INFN - Sezione di Roma - Piazzale Aldo Moro 2, 00185, Roma, Italy}
\newcommand{\MPP}{Max-Planck-Institut für Physik, D-85748 Garching, Germany}
\newcommand{\Sapienza}{Dipartimento di Fisica, Sapienza Università di Roma, P. le A. Moro 2, 00185 Roma, Italy}
\newcommand{\Grenoble}{Université Grenoble Alpes, CNRS, Grenoble INP, Institut Néel, 38000 Grenoble, France}
\newcommand{\Coimbra}{LIBPhys-UC, Departamento de Fisica, Universidade de Coimbra, P3004 516 Coimbra, Portugal}

\author{G.~Del~Castello~\orcidlink{0000-0001-7182-358X}}\email[Corresponding author: ]{giorgio.delcastello@roma1.infn.it}\affiliation{\INFNRoma}

\author{M.~Mancuso\orcidlink{0000-0001-9805-475X}}\affiliation{\MPP}
\author{A.~Langenk{\"a}mper}\affiliation{\MPP}
\author{A.~Bento}\affiliation{\MPP}\affiliation{\Coimbra}
\author{C.~Bonomo\orcidlink{0009-0001-0336-3857}}\affiliation{\INFNRoma}\affiliation{\Sapienza}
\author{M.~Cappelli\orcidlink{0009-0002-6148-5964}}\affiliation{\INFNRoma}\affiliation{\Sapienza}
\author{F.~Casadei}\affiliation{\MPP}
\author{E.~Cipelli}\affiliation{\MPP}
\author{A.~Cruciani\orcidlink{0000-0003-2247-8067}}\affiliation{\INFNRoma}
\author{D.~Delicato\orcidlink{0009-0005-0516-6872}}\affiliation{\Grenoble}
\author{F.~Dominsky}\affiliation{\MPP}
\author{M.~Folcarelli\orcidlink{0009-0009-7799-2515}}\affiliation{\INFNRoma}\affiliation{\Sapienza}
\author{D.~Hauff}\affiliation{\MPP}
\author{B.~Mauri}\affiliation{\MPP}
\author{C.~Moore}\affiliation{\MPP}
\author{L.~Pesce\orcidlink{0009-0001-5659-4691}}\affiliation{\INFNRoma}\affiliation{\Sapienza}
\author{F.~Petricca\orcidlink{0000-0002-6355-2545}}\affiliation{\MPP}
\author{F.~Pröbst}\affiliation{\MPP}
\author{D.~Quaranta\orcidlink{0009-0000-2954-4456}}\affiliation{\INFNRoma}\affiliation{\Sapienza}
\author{M.~Vignati\orcidlink{0000-0002-8945-1128}}\affiliation{\INFNRoma}\affiliation{\Sapienza}
\author{M.~Zanirato}\affiliation{\MPP}

\begin{abstract}
Lowering the energy threshold of phonon-mediated kinetic inductance detectors is a key step toward next-generation cryogenic calorimeter arrays with large target mass and high sensitivity. In this work, we demonstrate the feasibility of integrating thin tungsten films, so far used for transition-edge sensors, into the fabrication of Kinetic Inductance Detectors (KIDs). Here we present the results from the characterization of a single KID prototype composed of a tungsten-aluminum bilayer of respectively 15~nm and 35~nm thicknesses. The detector was successfully calibrated, and the superconducting film characteristics were measured by operating the device at different temperatures. We show that, by exploiting the proximity effect of superconductors, the device combines the relatively high critical temperature of aluminum with the large kinetic inductance contribution of tungsten. The main feature is a high kinetic inductance, which accounts for 36\% of the total inductance, combined with a superconducting critical temperature of around 630~mK. 

\end{abstract}

\maketitle

\paragraph*{\textbf{Introduction}}
\label{sec:intro}
In the past decade, phonon-mediated superconducting cryogenic detectors have been the state-of-the-art technology to detect small energy depositions produced by rare events~\cite{CRESST_2TES:2024,bullkid2024,tesseract2025,supercdms:2026,nucleus2025,Cosinus2024}. These devices are usually composed of a superconductive sensor coupled to a crystal absorber. The phonon mediation allows sensing of a large absorber mass, typically ranging from a fraction to tens of grams, with a small superconductive metal film. Moreover, the low energy cost to produce the signal carriers (phonons and quasiparticles) allows these detectors to achieve thresholds as low as 1~eV~\cite{1eV_tesseract}.

Currently, Transition-Edge Sensors (TESs) based on thin tungsten films~\cite{CRESST_2TES:2024,tesseract2025,nucleus2025,Cosinus2024,1eV_tesseract} are among of the best performing devices in terms of energy threshold and resolution. On the other hand, TESs suffer greatly from mass scaling issues, since their multiplexing is complicated, and currently no phonon-mediated experiment has demonstrated such a feature. For this reason, the use of phonon-mediated KIDs, like the one developed by the Calder, BULLKID and KIPM projects~\cite{calder17,bullkid2024,kipm}, has been investigated and new detector concepts are being proposed. While phonon-mediated KIDs have proven to be multiplexable~\cite{bullkid2026}, further developments are needed to improve their energy threshold to be on par with their TES counterpart.

Different strategies can be devised to improve the response of KIDs. One is to explore different detector geometries like the one proposed in Refs.~\cite{funnel,funnel2}. Another way is to employ different superconductors, as demonstrated in Refs.~\cite{calderaltial,TiN_Gao,HfKids}. Here, we investigate this second strategy by combining the high responsivity of thin tungsten films, typically used for TESs, with the aluminum films typical of KIDs. We realized a W-Al KID phonon-mediated detector, which, to the authors' knowledge, represents the first device of its kind.

\paragraph*{\textbf{Signal generation in a Kinetic Inductance Detector}}
\label{sec:respo}
A KID consists of a superconducting LC resonator~\cite{kid_design}, where the inductance $L$ is the sum of two terms: the magnetic inductance $L_g$ and the kinetic inductance $L_k$. An energy deposition $E$ in the crystal produces athermal phonons that are absorbed by the KID and can break Cooper-pairs, releasing quasiparticles. This process transiently reduces the Cooper-pair density, increasing the kinetic inductance $L_k$ and reducing the resonant frequency $f_r$~\cite{kid_design,mazin2005mkid,modello1,modello2}.

Usually, the transient change of the resonance of a KID is detected by monitoring the phase and magnitude modulation of the complex single pole transmission function $S_{21}(y)$ of the resonator~\cite{kid_germanio}:
\begin{equation}\label{eq:S21}
    S_{21}(y) = 1-\frac{Q}{Q_c} \frac{1}{1+2jy}
\end{equation}
 where $Q = (Q_c^{-1} + Q_i^{-1})^{-1}$, $Q_c$ is the coupling quality factor between the KID and the radio-frequency signal, and $Q_i$ is the internal quality factor~\cite{kid_design}, which quantifies the internal losses of the sensor. The parameter $y~=~Q\frac{f_r-f_\text{readout}}{f_r}$ describes the frequency difference between the readout tone the resonant frequencies in units of the resonance width $f_r/Q$. Usually, the phase readout $\varphi$ provides a better signal-to-noise ratio (SNR)~\cite{kid_design} with respect to the magnitude one and therefore, it is adopted in this work.

The signal phase responsivity of a KID is~\cite{kid_design, mazin2005mkid, kid_germanio, resp, temperature_scan}:
\begin{align}\label{eq:responsivity}
        \frac{d\varphi}{dE} =  \frac{\eta}{V_{\mathrm{KID}}} \frac{\alpha S_2(f_r, T)Q}{N_0\Delta_0^2} \frac{1}{1+4y^2}\approx \notag\\ \frac{\eta}{V_{\mathrm{KID}}} \cdot\frac{\alpha S_2(f_r, T)Q}{N_0\Delta_0^2}
\end{align}
where $\eta$ is the energy to pair-breaking conversion efficiency, $V_{\mathrm{KID}}$ is the active inductor volume, 
\begin{equation}\label{eq:alpha}
\alpha = \frac{L_k}{L_k+L_g}
\end{equation} 
is the kinetic inductance fraction, and $S_2(f_r, T)$ is a dimensionless factor of order one from Mattis-Bardeen theory. The approximation in Eq.~\ref{eq:responsivity} was done under the reasonable approximation that the readout tone and the resonant frequency of the device are very close and no correction needs to be applied. Finally, $N_0$ is the single-spin volume density of electron states at the Fermi energy and $\Delta_0 \simeq \frac{3.5}{2}k_BT_c$ is the superconducting energy gap, with $k_B$ the Boltzmann constant and $T_c$ the superconducting critical temperature. 

To understand how the superconducting film properties affect the responsivity of a KID, it is worth noting that the kinetic inductance $L_k$ of a superconducting film roughly scales as~\cite{Annunziata_2010}:

\begin{equation}
    L_k\propto L_k^\square \propto \frac{R^\square}{T_c}
\end{equation}
where $R^\square$ and $L_{k}^\square$ are, respectively, the sheet resistance in the non-superconducting state and the sheet kinetic inductance. In Eq.~\ref{eq:responsivity}, $\alpha$ and $\Delta_0$ are the parameters most strongly influenced by the choice of superconductor. Consequently, the responsivity of a KID scales approximately as:

\begin{equation}\label{eq:responsivity2}
        \frac{d\varphi}{dE} \propto \frac{R^\square}{T_c^3} 
\end{equation}

From the above equation it is clear that choosing a low $T_c$ superconductor, possibly with a high $R^\square$ could greatly enhance the responsivity of a KID. In this context, thin $\alpha$-tungsten films ($T_c\sim~15~\text{mK}$~\cite{CRESST_tes,CRESST_2TES:2024}) emerge as a compelling candidate to boost KID responsivity when compared to the standard aluminum film with $T_c\sim 1.2$~K, considering also that $R^\square_\text{W}/R^\square_\text{Al}\sim 4$ (see Refs.\cite{res_w,res_al}). On the other hand, to achieve optimal SNR conditions a KID must be operated at around $T_c/6$~\cite{Battistelli2015-ow}, since the minimum temperature reachable with a dry dilution refrigerator is around 10~mK, a $T_c\geq 60$~mK is required. For this reason in this study we produced a tungsten-aluminum bilayer, which depending on the relative film thicknesses would allow to tune the superconductive critical temperature from 15~mK to 1.2~K by exploiting the proximity effect of superconductors~\cite{proximity}.

\paragraph*{\textbf{Detector Description}}
\label{sec:detector}
\begin{figure}
    \centering
    \includegraphics[width=1.02\linewidth]{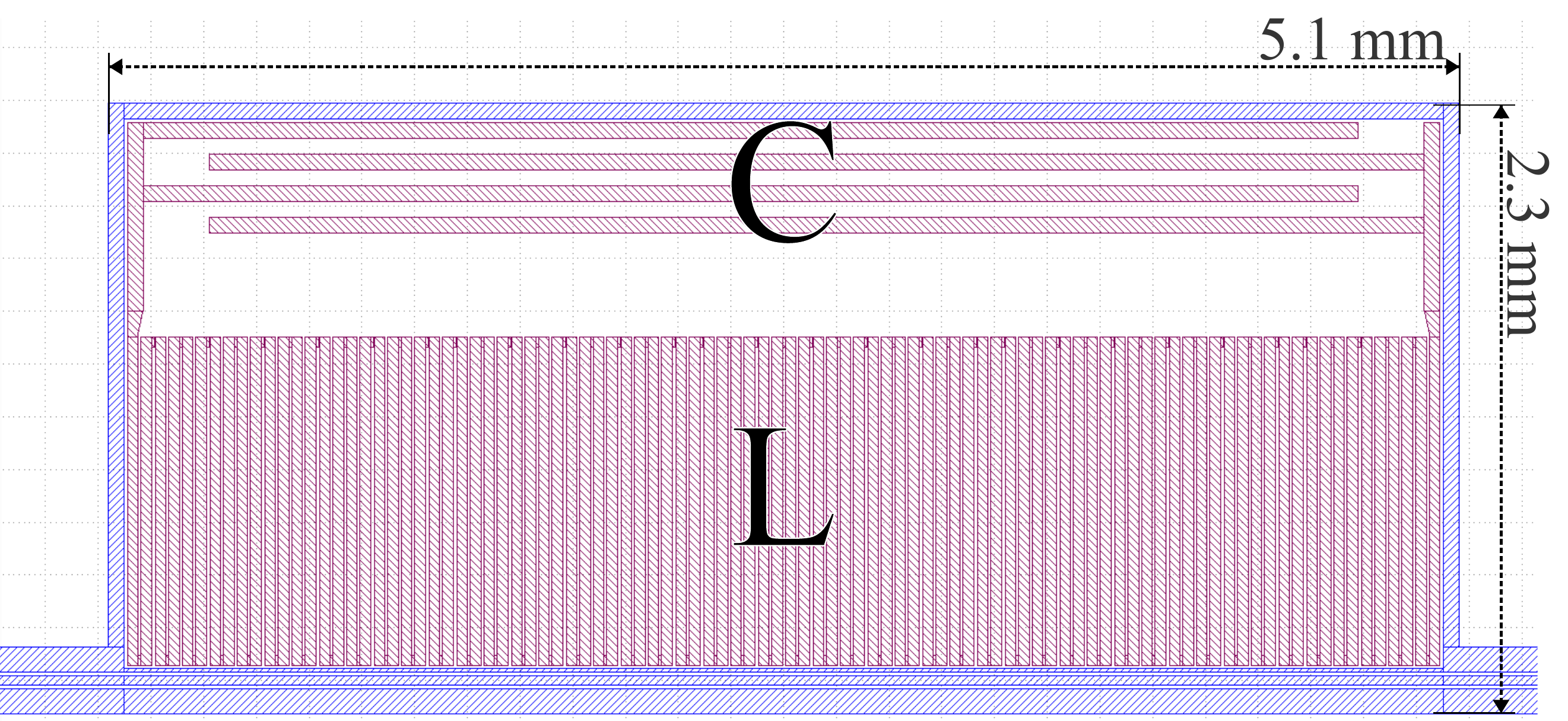}
    \includegraphics[width=1\linewidth]{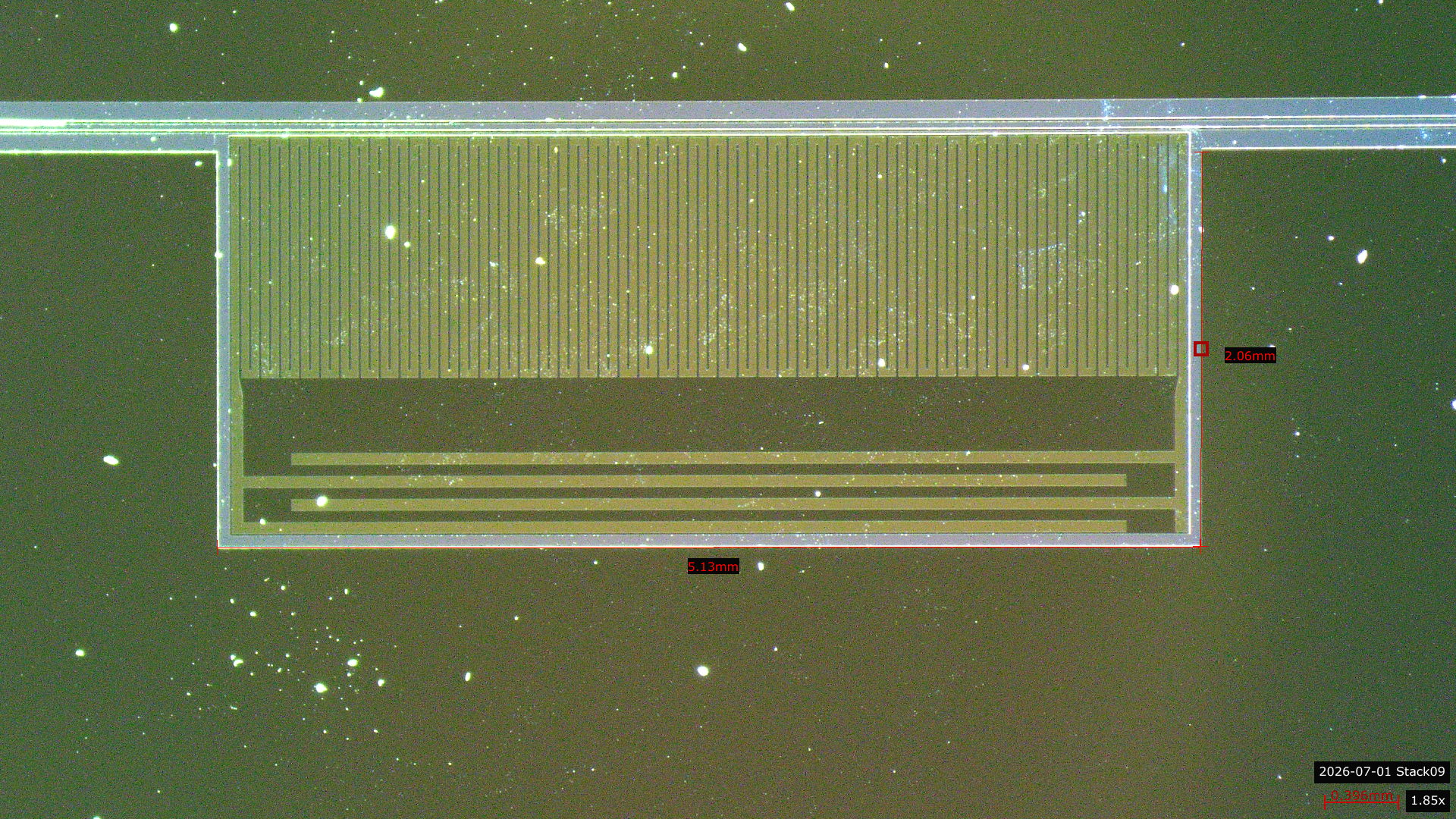}
    \caption{\textbf{Top}: Design of the sensor used in this work: in blue the feedline and guard ring are highlighted, while the KID is displayed in red. The letter C and L respectively highlight the capacitor and the inductor of the design. The blue and red areas indicate a different thickness of the metal films as presented in Table~\ref{tab:metal_thickness}. \textbf{Bottom}: Picture of the detector with its dimensions. In the picture it is visible how the different elements have different film thicknesses.}
    \label{fig:kid-design}
\end{figure}

The design of the KID used in this study, shown in Fig.~\ref{fig:kid-design}, has an inductive meander (L) area of 4.8~mm$^2$ and is inductively coupled with a coplanar waveguide feedline. Both these features make it similar to the Calder17 aluminum device discussed in Ref.~\cite{calder17}, whose main parameters are shown in Tab.~\ref{tab:calder17params}, which therefore will be used as comparison. The Calder17 resonator featured high $Q_i$ and $Q_c$ was designed in order to have high Q. On the other hand, the $Q_c$ of the W-Al KID was intentionally lowered to face a possible low $Q_i$ due to the new superconductors involved; the reason being that, as visible from Eq.~\ref{eq:S21}, it is advisable to have $Q\sim Q_c$.

\begin{table}
    \centering
    \caption{Main characteristics of the Calder17~\cite{calder17} aluminum device. Anticipating the results of this work, the table contains also all the relevant parameters for the W-Al KID for an easier comparison.}
    \label{tab:calder17params}

    \vspace{1mm}
    \begin{tabular}{c|c|c}
         & Calder17 & W-Al KID\\
         \hline
         \hline
         Active Area [mm$^2$] & 4.0& 4.8\\
         V$_\text{KID}$ [$10^{-6}$mm$^3$] & 240& 240\\
         V$_\text{Metal}$[$10^{-6}$mm$^3$] & 580& 1290\\
         Frequency [MHz] & $\sim 2600$& $715$ \\
         Q[k] &147& $8$\\
         Q$_c$[k] & 156& $11$ \\
         Q$_i$[k] & $>2\cdot 10^3$ & $31$\\
         $\alpha$ &$0.0254$ &$0.3568$\\
         \makecell{$\frac{d\varphi}{dE}\left[\frac{\text{mrad}}{\text{keV}}\right]$}&$5.8$& $5.9$ \\
         \makecell{Noise RMS [mrad]}&$0.61$& $1.02$ \\
    \end{tabular}
\end{table}

This resonator was simulated using SONNET~\cite{sonnet2025}, which allows to predict the resonant frequency and the coupling quality factor $Q_c$ of the resonator, provided the geometry of the KID and the estimation of $L_{k}^\square$. Since no reliable expectation for the kinetic inductance was available at the design stage, the first simulation was performed with this value set to zero. This approach provides an upper limit on the expected resonance frequency and accurately describes the coupling quality factor $Q_c$, while $Q_i$ is set to infinity. The corresponding results are reported in the first row of Tab.~\ref{tab:losses_simures}.

\begin{table}
    \centering
    \caption{SONNET simulation parameters and results. The $L_{k}^\square$ parameter is set as input to the simulation, while the resonant frequency and the quality factor are taken from analyzing the simulation results. The estimated statistical errors are negligible, while the systematic errors are evaluated by varying the sensor size of $\pm1\%$, which corresponds to the fabrication precision. The case with $L_{k}^\square = 0 \frac{\text{pH}}{\square}$ is used to estimate the resonator parameters in the absence of kinetic inductance, while the simulation with $L_{k}^\square = 4.33 \frac{\text{pH}}{\square}$ reproduces the measured resonant frequency, as described later in this work. For this reason, in the second simulation the errors are associated to the value of $L_{k}^\square$ and not to the resonant frequency.}
    \label{tab:losses_simures}

    \vspace{1mm}
    \begin{tabular}{c|c|c}
         $L_{k}^\square~\left[\frac{\text{pH}}{\square}\right]$  & Frequency [MHz] & $Q_c$ [k] \\
         \hline
         \hline
         0 & \makecell{$891.8\pm 6.8^\text{syst}$} & $7.4\pm1.0^\text{syst}$\\
         \makecell{$4.33\pm 0.15^\text{syst}$} & $715$ & $11.2 \pm 1.9^\text{syst}$\\
         \hline
         \multicolumn{3}{c}{\rule{0.95\linewidth}{0pt}}
    \end{tabular}
\end{table}

The KID was produced at the microfabrication facilities of the Max-Planck-Institut f{\"u}r Physik where the TESs for the CRESST and NUCLEUS experiments~\cite{CRESST_2TES:2024,nucleus2025} are also produced. While the Calder17 aluminum device was produced using an electron-gun evaporator, the W-Al KID presented here was produced by sputter deposition. Tungsten and aluminum were deposited in the same vacuum chamber, without breaking the vacuum between the two depositions. The full KID pattern was then defined by positive lithography and etching. The feedline and ground plane (blue elements in Fig.~\ref{fig:kid-design}, top) were subsequently thickened with a second aluminum deposition structured through a lift-off process. This last step, while not required for operating a KID, was done to reduce phonon absorption by the inhert structures through the phonon recycling phenomenon~\cite{recycling}. The final result of the production cycle yielded the resonator in Fig.~\ref{fig:kid-design} (bottom). Using a profilometer, it was also possible to measure the thickness of the deposited metal films which are reported in Tab.~\ref{tab:metal_thickness}. The thicknesses of the two metal layers were chosen to have $T_c\sim0.5$~K and the same active volume as Calder17.

\begin{table}
    \centering
    \caption{Metal thicknesses measured using a profilometer after detector fabrication. The colors next to each component refer to the colors in the top panel of Figure~\ref{fig:kid-design}.}
    \label{tab:metal_thickness}
    \vspace{1mm}
    \begin{tabular}{c|c|c}
         Component & W Thickness& Al Thickness\\
         \hline
         \hline
         KID (red) & 15 nm & 35 nm \\
         Feedline (blue) & 15 nm & 150 nm \\
         \hline
         \multicolumn{3}{c}{\rule{0.95\linewidth}{0pt}}
    \end{tabular}

\end{table}

\paragraph*{\textbf{Experimental Setup}}
\label{sec:setup}
 The device was operated in a dilution $^3\mathrm{He}/^4\mathrm{He}$ cryostat, at a base temperature $T_0~\simeq~30~\text{mK}$. During the operation two magnetic shields were placed around the detector. The first is a case made of Cryophy$^\text{\textregistered}$ thermalized at the same temperature as the detector, while the second, made of MuMETAL$^\text{\textregistered}$, was placed outside the cryostat. This is the usual setup employed when characterizing the Calder and BULLKID detector arrays~\cite{calder17,calderaltial,bullkid2024,bullkid2026}, which is able to attenuate the vertical component of earth's magnetic field of around a factor 100. This is required since the presence of a magnetic field, especially directed othogonally to the superconducting film, lowers the internal quality factor of the resonators.
 
The detector was mounted into a copper holder and was mechanically fixed in place by Teflon supports. The device was electrically coupled to the readout coaxial line, running through the cryostat from the outside down to the device, by aluminum wire bonding to the two $50\,\mathrm{\Omega}$ launchers. The readout line was equipped with 46~dB input attenuation, deployed at various temperature stages, to reduce the effective noise temperature of the system. The output of the line was amplified through an HEMT low-noise amplifier thermalized at the 4~K stage. The radio-frequency signal used to excite the KID was generated and read back at room temperature by a commercial Ettus X300 board~\cite{x300_specs} with custom firmware and software~\cite{minutolo}.

To calibrate the detectors, an optical fiber was placed facing the center of the crystal substrate on the side opposite to the lithography. The fiber was pulsed with $400\,\mathrm{nm}$ photon bursts generated at room temperature using the LEDs of the LANTERN calibration system~\cite{lantern,lantern2}.

\paragraph*{\textbf{Temperature Scan and Film Properties}}
\label{sec:tempscan}

At base temperature, corresponding to around 30~mK, the resonators' parameters are:
\begin{align*}
    &f_r^\text{30~mK} = (715.185\pm 0.010)~\text{MHz} \\
    &Q=8000\\
    &Q_c = 10700 \\
    &Q_i=31700
\end{align*}

where the reported error on $f_r^\text{30~mK}$ is the frequency spacing used for the transmission scan. By comparing the resonant frequency $f_r^\text{30~mK}$ with the simulation performed without kinetic inductance (see first row of Tab.~\ref{tab:losses_simures}) it is possible to estimate the kinetic inductance fraction $\alpha$ of Eq.~\ref{eq:alpha}:
\begin{equation}\label{eq:alpha}
    \alpha = 1-\left(\frac{f_r^{30~\text{mK}}}{f_r^{L_k=0}}\right)^2 = 0.3569 \pm0.0096^\text{syst}
\end{equation}
where $f_r^{L_k=0}=\left(891.8 \pm 6.8^\text{syst}\right)~\text{MHz}$ is the resonant frequency simulated assuming $L_{k}^\square = 0$. This value is more than an order of magnitude higher than the one reported for Calder17, amounting to $\alpha= 0.0254\pm 0.0028$, and such an increase is due to the presence of tungsten. By introducing a non-zero value of the kinetic inductance in the simulation to match the measured resonance frequency, one can estimate the sheet kinetic inductance to be $L_k^\square=~(4.33\pm 0.15^\text{syst})\frac{\text{pH}}{\square}$ (see simulation results in Tab.~\ref{tab:losses_simures}), which, in line with the previous result, is around an order of magnitude greater than the typical values for an aluminum film which is approximately $0.5\frac{\text{pH}}{\square}$~\cite{allk} .

\begin{figure}
    \centering
    \includegraphics[width=1\linewidth]{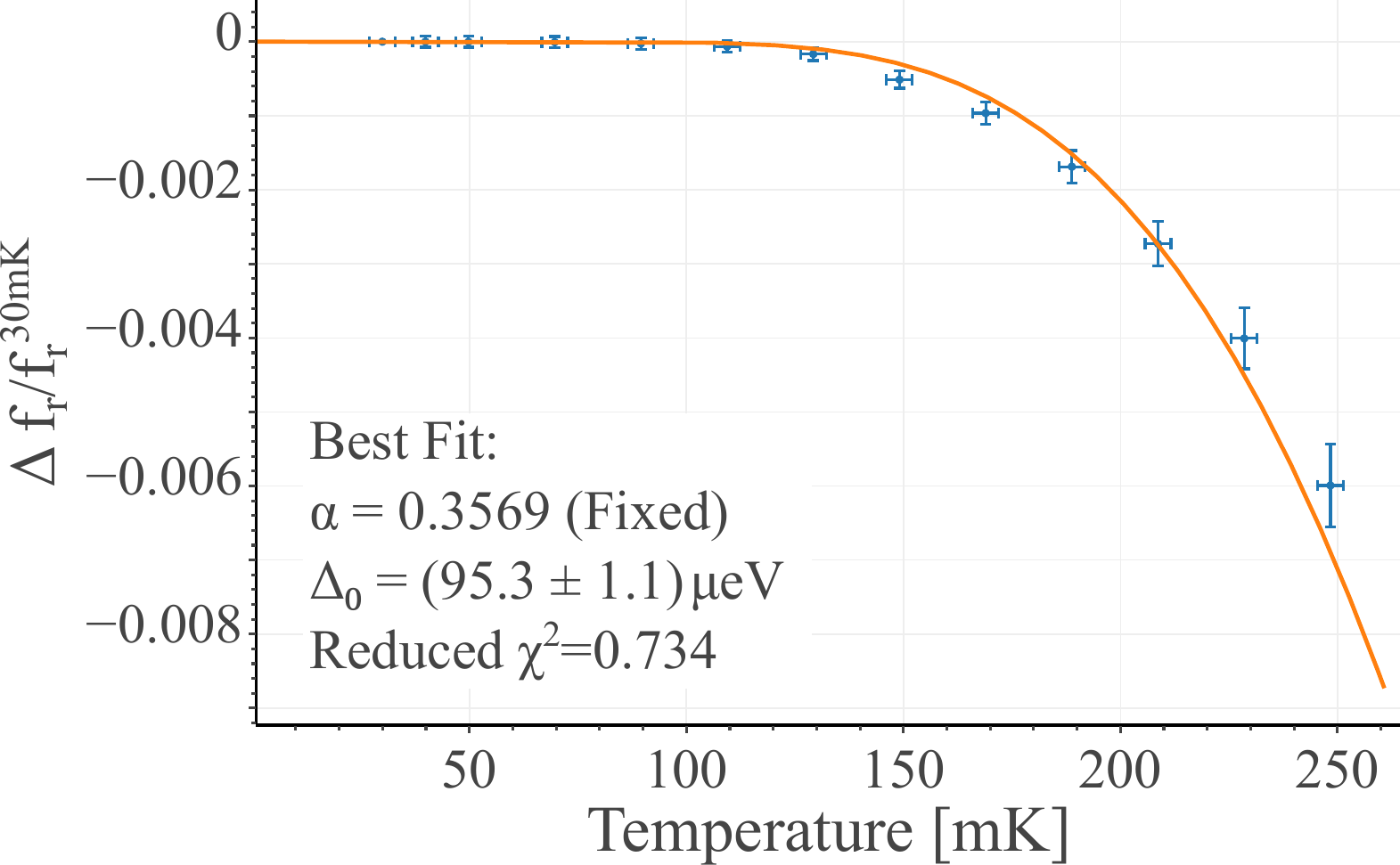}
    \caption{Shift of the resonant frequency with temperature. The data points are collected by measuring the resonance at each temperature stage. The plot shows the trend of fractional shift of the resonant frequency with respect to the one taken at the lowest temperature (30~mK). The points are fitted using the BCS prediction; the best-fit line and results are shown in the plot. The errors in the plot are given by the frequency resolution used during the measurement of the resonance, and a 3~mK uncertainty is used for the temperature scale.}
    \label{fig:tempscan}
\end{figure}

A second characterization consisted in performing a transmission measurement of the resonator operated at different temperatures. This temperature scan was performed using a heater placed at the coldest cryostat stage, and the detector was operated with a $-81$~dBm readout tone, a value sufficiently low to reduce self-heating effects. The measured quality factors and resonant frequencies from the scan can be used to fit for the Cooper-pair binding energy $\Delta_0$ (see Eq.~\ref{eq:responsivity}), using the BCS formalism~\cite{temperature_scan}. In Fig.~\ref{fig:tempscan}, the best-fit results are shown along with the measured data; in the fitting procedure, the value $\alpha$ was fixed to the previous estimation. Moreover, since $2\Delta_0~\simeq~3.5k_BT_c$, the fitted value shows a critical transition temperature of $T_c\simeq630~\text{mK}$, which, as expected, lies well below the 1.2~K critical temperature of aluminum. It is worth noting that the BCS theory might be limited in the case of inhomogeneous and multilayer superconductors. A more detailed threatment is beyond the scope of this work and can be found in Ref.\cite{Zhao_2018}.

\begin{figure}
    \centering
    \includegraphics[width=1\linewidth]{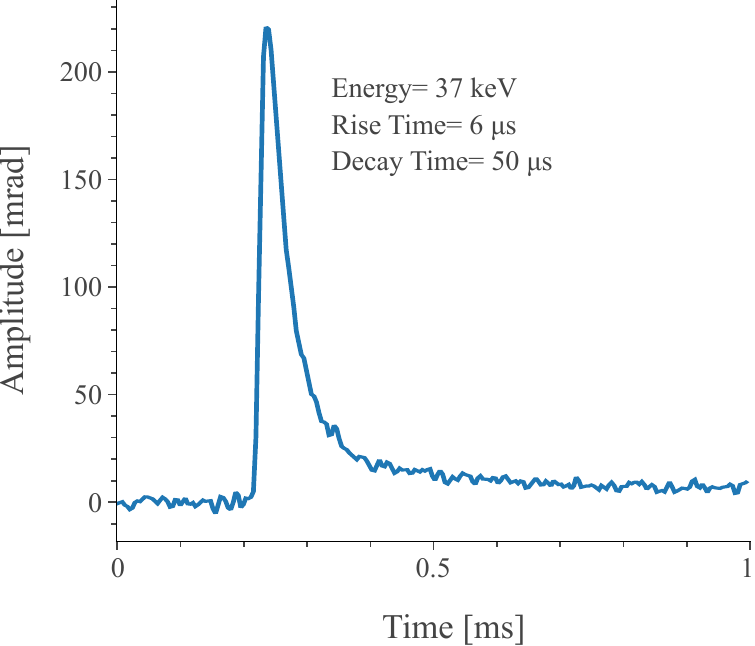}
    \caption{Pulse acquired during the detector calibration. The pulse reconstructed energy and its time constants are reported as well. The rise time of the pulse is a factor 2 higher than the resonator ring time $Q/\pi f_r\approx 3.~\mu\text{s}$, which can be due to the phonon lifetime in the substrate. The fast decay time instead points to a fast quasiparticle recombination time possibly linked to the low internal $Q_i$.}
    \label{fig:pulse}
\end{figure}

\paragraph*{\textbf{Calibration}}
\label{sec:meas}
Before performing the calibration of the detector, the signal-to-noise conditions were optimized both in terms of temperature and operating power. The optimization was done by measuring, for each working condition, the SNR of the reconstructed pulses in the KID following fixed energy depositions of photons from the calibration system. The highest SNR was found to be at a temperature of 110~mK with an operating power of $-66$~dBm. An example pulse acquired in the best SNR condition is shown in Fig.~\ref{fig:pulse} along with its time constants and energy.

The detector calibration was performed following the procedure described in Refs.~\cite{lantern,lantern2}, which consists of studying the broadening of the amplitude distributions produced by LED-induced pulses. A matched filtering algorithm was used to achieve the optimal SNR conditions, as described in Refs.~\cite{pydiana,bullkid_analysis}. The results of this calibration are:
\begin{equation*}
    \frac{d\varphi}{dE}= (5.90\pm0.21)\frac{\text{mrad}}{\text{keV}} \qquad \sigma_0=(172.2\pm6.4)~\text{eV}
\end{equation*}
where $\frac{d\varphi}{dE}$ is the responsivity of the device and $\sigma_0$ is the calibrated noise level after the application of the matched filter. The responsivity ratio between this detector and Calder17 is:
\begin{equation}\label{eq:meas_ratio}
\left.\dfrac{\left.\frac{d\varphi}{dE}\right|_\text{W-Al}}{\left.\frac{d\varphi}{dE}\right|_\text{Calder17}}\right|_\text{Measured} = \frac{5.9}{5.8}\sim 1
\end{equation}

The expected responsivity ratio can be obtained by substituting the parameters of Tab.~\ref{tab:calder17params} in Eq.~\ref{eq:responsivity} for both the W-Al KID and Calder17:
\begin{equation}\label{eq:resp_ratio}
\left.\dfrac{\left.\frac{d\varphi}{dE}\right|_\text{W-Al}}{\left.\frac{d\varphi}{dE}\right|_\text{Calder17}}\right|_\text{Expected} = \dfrac{\left(\frac{\eta\alpha S_2(f_r, T) Q}{N_0\Delta_0^2}\right)_\text{W-Al}}{\left(\frac{\eta\alpha S_2(f_r, T) Q}{N_0\Delta_0^2}\right)_\text{Calder17}}\sim 1.6
\end{equation}
where $N_0^\text{Calder17}= 1.72 \cdot 10^{10}\text{eV}^{-1} \mu\text{m}^{-3}$ and $N_0^\text{W-Al}= 1.46 \cdot 10^{10}\text{eV}^{-1} \mu\text{m}^{-3}$ were calculated by following Ref.~\cite{proximity,kittel}, while $S_2(f_r, T)^\text{Calder17}=3.02$ and $S_2(f_r, T)^\text{W-Al}=3.20$ and $\eta_\text{W-Al}/\eta_\text{Calder17}=0.45$. The contribution of $\eta_\text{W-Al}/\eta_\text{Calder17}$ to Eq.~\ref{eq:resp_ratio} is the one affected by the greatest uncertainty since the phonon transmission properties between silicon and tungsten are not known; assuming them similar to the Si-Al interface of Calder17, the ratio is then dominated by the phonon absorption competition between the active and the passive volumes of the sensors~\cite{martinez}, meaning that: 
$$\frac{\eta_\text{W-Al}}{\eta_\text{Calder17}}\sim\frac{\left(\frac{V_\text{KID}}{V_\text{Metal}}\right)_\text{W-Al}}{\left(\frac{V_\text{KID}}{V_\text{Metal}}\right)_\text{Calder17}}= 0.45$$

where $V_\text{Metal}$ is the sum of the active volume $V_\text{KID}$ and the remaining metalized volume of the passive structures of the sensor. 

The discrepancy between the measured value of 1 and the expected value of 1.6 of the responsivity ratio then points to a lower phonon transmission between Si and W with respect to the Si-Al case. It is possible to roughly estimate the phonon transmission in the two cases by using the acoustic wave propagation formalism described in Ref.~\cite{earthquake}, which gives a 1.9 times smaller transmission for Si-W than for the case of Si-Al. This shifts the expectation of the responsivity ratio of Eq.~\ref{eq:resp_ratio} to 0.8, in line with the measured value.

\paragraph*{\textbf{Conclusions and Outlook}}
\label{sec:conclusions}
This work presents the first realization of a kinetic inductance detector integrating tungsten thin-film technology typical of transition-edge sensors. The sensor, based on a sputtered aluminum-tungsten bilayer, shows a superconductive band gap energy of about 95~$\mu$eV, corresponding to a superconducting critical temperature of 630~mK. The device was successfully fabricated and operated in a dry dilution cryostat, achieving a high kinetic inductance fraction of 36\%, about a factor of 10 above typical aluminum values. These results highlight the strong potential of this approach for highly responsive kinetic inductance detectors. The measured responsivity of the W-Al KID is $5.9~\frac{\text{mrad}}{\text{keV}}$ and is currently limited by the low resonator quality factor. While a low coupling quality factor was chosen by design, the internal quality factor remains below that of aluminum KIDs such as Calder17, possibly due to the sputtering procedure of the	 superconducting films. As visible from Eq.~\ref{eq:resp_ratio}, reaching Calder17-like quality factors is expected to increase the responsivity up to a factor of 20, mostly due to the $\alpha$ and $\Delta_0$ contributions.

Several clear paths can further improve this proof of concept device. The first is to raise the resonator quality factor through fabrication optimization. In addition, tuning the aluminum and tungsten thicknesses can better optimize the critical temperature. The phonon collection efficiency is still to be measured, and improving the interface metal between tungsten and the crystal substrate, possibly turning to a Al-W-Al trilayer, should enhance phonon transmission. Finally, more advanced KID geometries and dedicated phonon-collecting structures~\cite{funnel} can further improve responsivity and noise performance. Taken together, these results provide strong motivation to pursue follow-up studies and fully exploit the potential of this technology.

\paragraph*{\textbf{Acknowledgments}}
\label{sec:acknowledgments}
This work was supported by the INFN, Sapienza University of Rome, the Max-Planck-Institut für Physik, and co-funded by the Deutsche Forschungsgemeinschaft (DFG, German Research Foundation) through the Sonderforschungsbereich (Collaborative Research Center) SFB1258 and by the European Union (ERC, DANAE, Grant No. 101087663).
Views and opinions expressed are however those of the authors only and do not necessarily reflect those of the European Union or the European Research Council. Neither the European Union nor the granting authority can be held responsible for them. We thank A. L. De Santis for the useful discussions on the phonon transmission properties.
\bibdata{references}
\bibliography{references}

\end{document}